\documentclass[12pt]{article}
\usepackage[dvips]{graphicx}

\def\PR#1#2#3{Phys. Rev. {\bf #1}, #2 (#3)}
\def\PRL#1#2#3{Phys. Rev. Lett. {\bf #1}, #2 (#3)}
\def\PL#1#2#3{Phys. Lett. {\bf #1}, #2 (#3)}

\def\NP#1#2#3{Nucl. Phys. {\bf #1}, #2 (#3)}

\def\PTP#1#2#3{Prog. Theor. Phys. {\bf #1}, #2 (#3)}
\def\EPJ#1#2#3{Eur.Phys.J. {\bf #1}, #2 (#3)}
\begin{document}
%
\title{The effect of Majorana phase in degenerate neutrinos}
\author{
{N. Haba$^{1}$}\thanks{E-mail address:haba@eken.phys.nagoya-u.ac.jp} 
{, Y. Matsui$^2$}\thanks{E-mail address:matsui@eken.phys.nagoya-u.ac.jp} 
{, N. Okamura$^3$}\thanks{E-mail address:naotoshi.okamura@kek.jp} 
{ and M. Sugiura$^2$}\thanks{E-mail address:sugiura@eken.phys.nagoya-u.ac.jp} 
\\
\\
\\
{\small \it $^1$Faculty of Engineering, Mie University,}
{\small \it Tsu Mie 514-8507, Japan}\\
{\small \it $^2$Department of Physics, Nagoya University,}
{\small \it Nagoya, 464-8602, Japan}\\
{\small \it $^3$Theory Group, KEK, Tsukuba Ibaraki 305-0801, Japan}}
\date{}
\maketitle
\vspace{-10.5cm}
\begin{flushright}
hep-ph/9908429\\
DPNU-99-25\\
KEK-TH-636
\end{flushright}
\vspace{10.5cm}
\vspace{-2.5cm}
%
%
\begin{abstract}

There are physical Majorana phases in the lepton
 flavor mixing matrix when neutrinos are Majorana
 fermions.
In the case of two degenerate neutrinos, 
 the physical Majorana phase plays the crucial role for
 the stability of the maximal flavor mixing between the second
 and the third generations 
 against quantum corrections. 
The physical Majorana phase of $\pi$ 
 guarantees the maximal mixing 
 to be stable against quantum corrections, 
 while the Majorana phase of zero 
 lets the maximal mixing be spoiled
 by quantum corrections when neutrino masses are of $O$(eV). 
The continuous change of the Majorana phase 
 from $\pi$ to $0$ makes the maximal mixing
 be spoiled by quantum corrections 
 with $O$(eV) degenerate neutrino masses.
On the other hand, 
 when there is the large mass hierarchy
 between neutrinos,  
 the maximal flavor mixing 
 is not spoiled by quantum corrections 
 independently of the
 Majorana phase. 

\end{abstract}


\newpage

%
%
\section{Introduction}

Recent neutrino oscillation experiments suggest
 the strong evidences of the tiny neutrino masses 
 and lepton flavor mixings\cite{solar4}-\cite{SK4}.
Studies of the lepton flavor mixing matrix,
 which is so-called Maki-Nakagawa-Sakata(MNS) matrix\cite{MNS},
 will give us important cues of the physics
 beyond the standard model.
One of the most important studies of the 
 lepton flavor mixing is 
 the analysis of the quantum correction on 
 the MNS matrix\cite{up2now}-\cite{HO1}.

In this paper we analyze 
 the effect of the Majorana phase for 
 the stability against quantum corrections 
 of the maximal mixing between the second and the
 third generations, 
 which is suggested by the atmospheric
 neutrino experiments\cite{Atm4,SK4}. 
There is one physical Majorana phase in the $2 \times 2$
 MNS matrix when neutrinos are Majorana
 fermions.
This Majorana phase 
 plays the crucial role for
 the stability against quantum corrections  
 when two neutrino masses are degenerate. 
When Majorana phase is equal to 
 zero, the maximal
 mixing is spoiled by quantum corrections 
 when neutrino masses are 
 of $O$(eV)\cite{HOS1}. 
On the other hand, the 
 maximal mixing is not spoiled by quantum corrections 
 when Majorana phase is 
 equal to $\pi$ 
 because the neutrino mass matrix has 
 the pseudo-Dirac texture\cite{SD1}. 
The continuous change of the Majorana phase 
 from $\pi$ to $0$ makes the maximal mixing
 be spoiled by quantum corrections when 
 neutrino masses are of $O$(eV).

When there is
 the large mass hierarchy between neutrinos,
 the lepton flavor
 mixing is stable against quantum corrections
 independently of the Majorana phase. 

\section{Stability of the maximal mixing and Majorana phase}

The neutrino mass matrix of the
 second and the third generations
\begin{equation}
\kappa = \left( 
\begin{array}{cc}
\kappa_{22}    & \kappa_{23} \\
\kappa_{23}    & \kappa_{33}
\end{array}
\right)
\end{equation}
\vspace{12pt}
is diagonalized as
\begin{equation}
 U^{T} \kappa \ U = D_{\kappa}\,,
\label{d-kappa}
\end{equation}
where $D_{\kappa}$ is given by 
\begin{equation}
 D_{\kappa} =
\left(
\begin{array}{cc}
 m_2 & 0 \\
   0 & m_3
\end{array}
\right)\,,
\label{mi}
\end{equation}
with $m_i \geq 0$ $(i=2,3)$. 
The unitary matrix $U$ 
 is defined as
\begin{equation}
\label{MNS}
U = 
\left(
\begin{array}{cc}
 \cos \theta_{23} & \sin \theta_{23} \\
-\sin \theta_{23} & \cos \theta_{23}
\end{array}
\right)
\left(
\begin{array}{cc}
1   & 0         \\
0   & e^{i \phi/2}
\end{array}
\right) \,,
\end{equation}
where $\theta_{23}$ is the mixing angle between the second 
 and the third generations, and
 $\phi$ denotes the physical Majorana phase of neutrinos.
In the diagonal base of charged lepton masses, 
 $U$ is just the MNS matrix.

Since the atmospheric 
 neutrino experiments suggest\cite{Atm4,SK4}
\begin{equation}
\sin^2 2 \theta_{23} \simeq 1 \,,
\end{equation}
we analyze the stability of the mixing angle 
 $\theta_{23}$ against
 quantum corrections around $\theta_{23}= \pi/4$. 
In this case, eq.(\ref{d-kappa}) shows 
\begin{eqnarray}
\kappa &=& U^{\ast} D_{\kappa} U^{\dagger} \nonumber \\
       &=& \displaystyle{\frac{1}{2}}
\left(
\begin{array}{cc}
 m_2 + m_3 e^{-i\phi}  & -m_2 + m_3 e^{-i \phi}\\
-m_2 + m_3 e^{-i \phi} &  m_2 + m_3 e^{-i\phi} 
\end{array}
\right)\,.
\label{eqn.kappa}
\end{eqnarray}
Quantum corrections change the form 
 of $\kappa$\cite{EL1,HMOS1} as  
\begin{eqnarray}
\label{k'}
\kappa' &=& 
\left(
\begin{array}{cc}
 1    &  0 \\
 0    & 1+ \epsilon 
\end{array}
\right) \kappa
\left(
\begin{array}{cc}
 1    &  0 \\
 0    & 1+ \epsilon 
\end{array}
\right)\,, \nonumber \\
 &=& 
\displaystyle{\frac{1}{2}}
\left(
\begin{array}{cc}
 m_2 + m_3 e^{-i\phi}  & (-m_2 + m_3 e^{-i \phi}) (1+\epsilon)\\
(-m_2 + m_3 e^{-i \phi})(1+\epsilon) &  (m_2 + m_3 e^{-i\phi})(1+2\epsilon) 
\end{array}
\right)+O(\epsilon^2)\,.
\end{eqnarray}
The unitary matrix $U'$ which diagonalizes
 $\kappa'$ shows us
 whether the maximal mixing between the second
 and the third generations is spoiled by 
 quantum corrections or not. 
The mixing angle ${\hat{\theta}_{23}}$ 
 which diagonalizes $\kappa'$ in 
 eq.(\ref{k'}) is given by 
\begin{equation}
\tan 2 \hat{\theta}_{23} = 
\displaystyle{\frac{1}{\epsilon}}\;
\displaystyle\frac{\delta m^2 }
{m_2^2 + m_3^2 + 2 m_2 m_3 \cos \phi}\;
+ \; O(\epsilon^0) \; ,
\label{mixing}
\end{equation} 
where 
\begin{equation}
\delta m^2 \equiv m_3^2 - m_2^2 \,,
\end{equation}
which is determined by the atmospheric 
 neutrino experiments\cite{Atm4,SK4}. 
Equation (\ref{mixing}) shows that 
 the mixing angle $\hat{\theta}_{23}$ 
 is stable (unstable) against quantum
 corrections when 
 the value of 
 $\delta m^2/(m_2^2 + m_3^2 + 2 m_2 m_3 \cos \hat{\phi})$
 is larger (smaller) than that of 
 $\epsilon$.

When two mass eigenvalues are degenerate as $m_2 \simeq m_3$,
 we can easily see the following facts at $\phi=0$, and $\pi$:
\begin{enumerate}
\itemsep 0em
\renewcommand{\labelenumi}{({\it\roman{enumi}})}
\item  When $\phi=0$,
 eq.(\ref{mixing}) derives 
\begin{equation}
 \tan 2 \hat{\theta}_{23}
 \simeq {1 \over \epsilon}\; {m_3 - m_2 \over m_3 + m_2}\; .
\end{equation}
This means that the mixing angle $\hat{\theta}_{23}$ is
 stable (unstable) when the value of $(m_3 - m_2)/(m_3 + m_2)$ is 
 larger (smaller) than that of $\epsilon$. 
When $m_3=O(1)$ eV, the value of $(m_3 - m_2)/(m_3+ m_2)$ 
 is small enough for the maximal mixing to be spoiled 
 by quantum corrections\cite{HOS1}.

\item  When $\phi= \pi$,  
 eq.(\ref{mixing}) derives 
\begin{equation}
 \tan 2 \hat{\theta}_{23}
 \simeq {1 \over \epsilon}\;{m_3 + m_2 \over m_3 - m_2}\; .
\label{11}
\end{equation}
This means that the maximal mixing is
 not spoiled by quantum corrections. 
It is because the absolute 
 values of off-diagonal elements in 
 $\kappa'$ are 
 much larger than those of diagonal elements, 
 which is so-called the pseudo-Dirac texture.
Equation (\ref{11}) is induced just from the initial
 value of $\theta_{23}= \pi /4$. 
For the general values of $\theta_{23}$, 
 it is satisfied that 
 $\tan 2 \hat{\theta}_{23}= 
 \tan 2 \theta_{23} +O( \epsilon^2)$\cite{HO1}.
This means the mixing angle of 
 $\hat{\theta}_{23}$ receives little changes from 
 quantum corrections for the general 
 values of $\theta_{23}$.
\end{enumerate}
The behaviors against quantum corrections 
 in cases of $(i)$ and $(ii)$ are completely
 different from each other as shown above. 
The physical Majorana phase of $\pi$ 
 guarantees the maximal flavor mixing 
 to be stable against quantum corrections, 
 while the Majorana phase of zero 
 lets the maximal mixing be spoiled
 by quantum corrections when $m_3=O\mbox{(eV)}$. 
However eq.(\ref{MNS}) suggests that they
 are connected with each other by the continuous change
 of Majorana phase $\phi$. 
Therefore this Majorana phase $\phi$ plays the crucial
 role for the stability of the lepton flavor 
 mixing angle against quantum corrections.

Figure 1 shows the contour plot of 
 $\sin^2 2 \hat{\theta}_{23}$ for the continuous changes of
 $\phi$ and $m_2{}^2$. 
We use the value of 
$ \delta m^2 = 3 \times 10^{-3} \mbox{eV}^2 $
\cite{Atm4,SK4}.
The value of $\epsilon$ is determined by two parameters of 
 $\tan \beta = 20$ and
 the intermediate scale $M_R=10^{13}\; \mbox{GeV}$\cite{HO1}. 
We input the low-energy data to $\theta_{23}$ 
 and $\delta m^2$, and show the value of $\hat{\theta}_{23}$
 at $M_R=10^{13}\; \mbox{GeV}$ in Fig.1.
The continuous change of the Majorana phase 
 from $\pi$ to $0$ makes the maximal mixing
 between the second and the third generations 
 be spoiled by quantum corrections when 
 $m_2^2 =O(1)$eV$^2$. 
The larger the value of $m_2{}^2$ becomes,
 the smaller the angle $\hat{\theta}_{23}$ 
 becomes. 
As we have shown at $(ii)$,
 the maximal mixing is not spoiled by
 quantum corrections around $\phi = \pi$ 
 independently of the value of $m_2{}^2$.

Figure 1 also shows that 
 the maximal mixing is stable against quantum corrections 
 around $m_2{}^2 \simeq 0$ 
 independently of the value of $\phi$. 
It is because 
 eq.(\ref{mixing}) derives 
\begin{equation}
 \tan 2 \hat{\theta}_{23}\simeq {1 \over \epsilon}
\label{12}
\end{equation}
when $m_2{}^2=0$. 
The physical Majorana phase can be rotated out
 by the field redefinition in this case. 
Equation (\ref{12}) suggests 
 that the maximal mixing is not spoiled by 
 quantum corrections 
 independently of the Majorana phase $\phi$ 
 when there is
 the large mass hierarchy between $m_2$ and $m_3$.
Equation (\ref{12}) is induced just from the initial
 value of $\theta_{23}= \pi /4$, and 
 for the general values of $\theta_{23}$, 
 it is satisfied that 
 $\tan 2 \hat{\theta}_{23}= 
 \tan 2 \theta_{23}(1 - \epsilon \:{\rm sec}2 \theta_{23}) +
 O(\epsilon^2)$\cite{HO1}. 
This means the mixing angle of 
 $\hat{\theta}_{23}$ receives little changes from 
 quantum corrections for the general 
 values of $\theta_{23}$.

This result implies that the 
 flavor 
 mixing matrix with the large mass hierarchy of 
 $|m_3| \gg |m_2| \gg |m_1|$
 is also stable against quantum corrections 
 in the three generation neutrinos. 
It has been shown by the numerical 
 analyses in Ref.\cite{HO1}.

\begin{figure}[t]
\begin{center}
\includegraphics[width=0.6\textwidth]{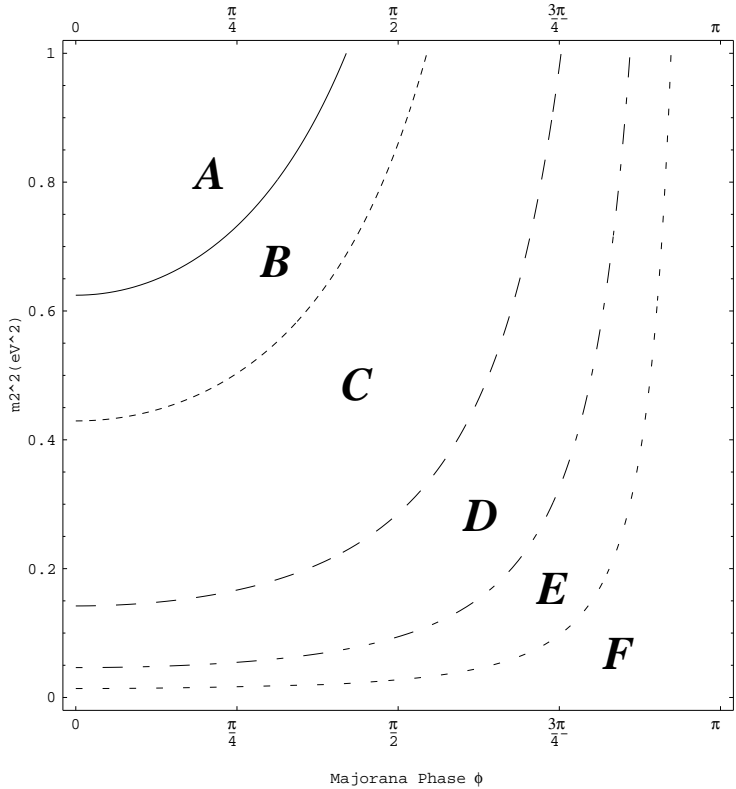}
\caption{%
 The contour plot of 
 $\sin^2 2 \hat{\theta}_{23}$ for the continuous changes of
 the Majorana phase $\phi$ and $m_2{}^2$ 
 (A: $\sin^2 2 \hat{\theta}_{23} < 0.05$,
  B: $ 0.05 \leq \sin^2 2 \hat{\theta}_{23} < 0.1$,
  C: $0.1 \leq \sin^2 2 \hat{\theta}_{23} < 0.5$,
  D: $0.5 \leq \sin^2 2 \hat{\theta}_{23} < 0.9$,
  E: $0.9  \leq \sin^2 2 \hat{\theta}_{23} < 0.99$,
  F: $0.99 \leq \sin^2 2 \hat{\theta}_{23}$). 
We use the experimental value 
 of $\delta m^2 = 3 \times 10^{-3}$ eV$^2$\cite{Atm4,SK4}.
The value of $\epsilon$ is determined by two parameters 
 of $\tan \beta = 20$ and
 $M_R=10^{13}$ GeV \cite{HO1}.
$\hat{\theta}_{23}$ is the mixing at 
 $M_R=10^{13}$ GeV.
}
\label{fig1}
\end{center}
\end{figure}

\section{Summary}

There are physical Majorana phases in the MNS matrix
 when neutrinos are Majorana fermions. 
The Majorana phase in the MNS matrix 
 plays the crucial role for
 the stability 
 of the maximal mixing between the second and the
 third generations against quantum corrections 
 when two neutrino masses are degenerate. 
 When the Majorana phase is equal to  zero
 and neutrino masses are of $O$(eV),
 the maximal mixing is spoiled by quantum corrections.
On the other hand, 
 when Majorana phase is 
 equal to $\pi$, 
 the 
 maximal mixing is
 not spoiled by the quantum corrections 
 because of the pseudo-Dirac texture of
 the neutrino mass matrix. 
The continuous change of the Majorana phase 
 from $\pi$ to $0$ makes the maximal mixing
 be spoiled by quantum corrections when 
 neutrino masses are of $O$(eV).

When there is
 the large mass hierarchy between neutrinos,
 the maximal mixing 
 is not spoiled by quantum corrections
 independently of the Majorana phase. 
This result can explain that the mixing angles 
 are stable against quantum corrections in 
 the three generation neutrinos 
 with the large mass hierarchies of 
 $|m_3| \gg |m_2| \gg |m_1|$.

We can also analyze the effects of neutrino 
 Majorana phases in 
 the cases of three generation neutrinos with mass
 hierarchies of $|m_2| \sim |m_1| \gg |m_3|$ 
 and  $|m_3| \sim |m_2| \sim |m_1|$ \cite{New}.

\section*{Acknowledgment}
NH and NO would like to thank M. Tanimoto for 
 helpful comments, and thank Summer Institute '99
 at Yamanashi, Japan for the kind hospitality. 
The work of NO is supported by the JSPS Research Fellowship
for Young Scientists, No.2996.

%
%

\end{document}